# Bluetooth sensors in phyphox with Arduino and MicroPython – Paving the way from an idea to an experiment for teachers and learners

Sebastian Staacks[1], Dominik Dorsel[1], Alexander Krampe[1], Marcel Hagedorn[1], Edward Leier[1], Heidrun Heinke[1], Christoph Stampfer[1]

[1]Institute of Physics I and II, RWTH Aachen University, Aachen, Germany

E-mail: staacks@physik.rwth-aachen.de

## Abstract

In order to extend the available sensors of smartphone experiments with cheap microcontroller-based external sensors, the smartphone experimentation app "phyphox" has been extended with a generic Bluetooth Low Energy interface. Since its application requires an in-depth understanding of the underlying technologies, the direct use of that interface for educational purposes is limited. To avoid this difficulty, the functionality was encapsulated into an Arduino and MicroPython library. With these, also educators and learners with only rudimentary programming knowledge can integrate an app-based interface into microcontroller projects with only few lines of code. This opens a wide range of new learning opportunities, which are described exemplarily.

## 1. Introduction

Experimentation plays an essential role in science education. This can be a simple demonstration by a teacher, some guided experimentation by students in class, an advanced lab course, a science project with much decision left to the students or the discovery of an everyday life phenomenon in informal education scenarios like in science museums. One of the benefits of educational experimentation is that students can get an insight to the scientific process including the planning of own experiments. In modern science this includes digital data acquisition, which is seen as an important competency to be acquired by students (Lahme et al., 2023). Designing labcourses with the goal to develop students' experimentation skills has also been demonstrated to improve their critical thinking (Walsh et al., 2022).

However, access to digital data acquisition can be limited by insufficient financial resources or limited staff to maintain and organize professional measurement equipment. These limitations have been overcome to some extent in recent years by two approaches. These are smartphone experiments on the one hand (Vogt et al., 2011; Pendrill & Rohlén, 2011; Chevrier et al., 2013; Vieyra et al., 2015; Monteiro & Marti, 2022) and cheap custom made sensors on the other, which have become easily available for educational purposes (Galeriu, 2013; Kubínová & Šlégr, 2015; Kuan et al., 2016; Kondaveeti et al., 2021). The former makes use of the vast collection of sensors in existing smartphones or tablets by using various apps to access, analyze and export sensor data. The latter is closely connected with the maker culture, where typically self-taught enthusiasts employ easy to use microcontroller platforms like Arduino or MicroPython and create also custom built measurement devices from a massive assortment of available sensors. This typically requires only cheap components, but at least basic electronics and programming skills.

Both approaches have limitations (Fig. 1). Smartphone experiments are limited to sensors included in the phone. This typically encompasses classical mechanics (accelerometer, gyroscope, microphone, pressure sensor) and some aspects of electromagnetism (magnetometer) and rudimentary optics (light sensor). Other areas of physics and even more so chemistry or biology remain inaccessible. On the other hand, microcontrollers usually lack in user-friendliness. Reading a cheap pH sensor or using an analog input for voltage measurements can be done in few lines of code. However, plotting the resulting data either involves export to additional software like Excel not allowing for real-time plotting of measured data or it demands elaborate coding and additional libraries and thus easily exceeds the competencies or the time budget of many potential users.

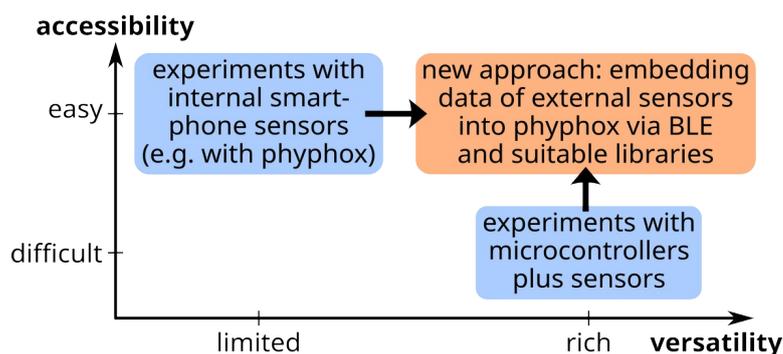

**Fig 1** The presented approach aims to combine the high versatility of microcontroller-based experiments with the high accessibility of smartphone-based experiments.

In this paper we present a Bluetooth interface for the well-established smartphone experimentation app "phyphox" (Staacks et al., 2018; Stampfer et al. 2020), which allows the combination of both approaches for powerful, versatile and still low-cost digital data acquisition (Fig. 1) in science education. Embedding phyphox into an Arduino or MicroPython project is done with an open library and only requires as little as three lines of code to plot sensor data over time. This makes digital data acquisition available for all science disciplines and for users among both educators and learners with limited programming experience.

## 2. Outline

As this paper explicitly targets also educators without in-depth programming knowledge, we will first introduce basic concepts of Bluetooth Low Energy (BLE). We will then describe the generic BLE support in phyphox, which allows easy use of existing sensors but requires deeper knowledge to extend to additional sensors. After discussing the hurdles to novices that are presented by this generic interface, we will introduce the Arduino and MicroPython libraries that allow easy incorporation of custom sensors for everybody.

Readers who are interested in using the Arduino or MicroPython libraries should also refer to the documentation and examples on the respective web pages[1].

---

1     Arduino: https://github.com/phyphox/phyphox-arduino
       MicroPython: https://github.com/phyphox/phyphox-micropython

## 3. Bluetooth Low Energy

Most modern battery-powered peripheral devices use Bluetooth Low Energy to communicate with smartphones. This choice often follows by ruling out the alternatives. Wired connections are often not desirable and the currently available Wifi standards require an existing infrastructure or using them would interfere with the connectivity of the phone. Most phones do not yet support connecting to the peripheral and a local Wifi network simultaneously, so the peripheral network would typically have to be registered into the local network instead. However, it should be noted that phyphox already features a network interface for use cases in which such infrastructure is available and advantageous. Examples for practical use in collecting data from many experiments on a central server have been described by Staacks et al. (2022).

Bluetooth Low Energy is a part of the Bluetooth standard along with "classical Bluetooth" since version 4.0. However, classical Bluetooth is a rather broad standard allowing for arbitrary custom communication protocols that cannot be realistically universally supported by an interface in phyphox. An additional hurdle to classical Bluetooth is that Apple devices only support peripherals if they carry the license "Made for iPhone" which is typically not the case for DIY electronics.

In contrast, Bluetooth Low Energy is supported by every modern Smartphone since Android 4.3 and iOS 5 and it follows a well-defined protocol which is aimed at battery-powered peripherals sending small data packets to a central device. It does not entirely prevent the use of a very specific custom protocol, but usually these are not required and communication even of proprietary devices is not too hard to understand without documentation.

The communication between two BLE devices is defined by the so-called Generic Attribute Profile (GATT). In typical sensor applications the phone is called the "central device" and may connect to several peripherals (for example different sensors), but instead of the phone acting as the server, each peripheral runs its own GATT server. Every value that is written to or read from the GATT server is structured into so-called "characteristics". For example, a climate sensor can have a characteristic for temperature, another one for humidity and one for $CO_2$ concentration. The characteristics are then structured into "services" and each characteristic and each service is identified by a Universally Unique Identifier (UUID). Instead of periodically reading new values from the server on the peripheral, the central device (i.e. the smartphone) subscribes to the characteristics it is interested in. The sensor will then subsequently send notifications to the central device when a new value is available.

The final essential concept of BLE is advertisement and typically happens before a central device connects to the server of a peripheral. This is how peripherals are found when scanning for devices in a connection dialog on the phone. The peripherals periodically broadcast small data packages with little information about themselves, which typically includes a short name and/or UUIDs of services that it provides.

## 4. Generic BLE support in phyphox

At first glance, phyphox is perceived by most users as a tool that simply plots a variety of sensor data over time and that has a list of predefined experiment configurations with specific data analysis and visualization. What most users do not notice at first is that such experiment configurations are not hard-coded, but described in a phyphox specific file format with XML syntax. This format defines data sources such as sensors, a sequence of data analysis steps, and visualization elements (i.e. graphs, simple values, but also interactive elements like buttons or value inputs).

The BLE support in phyphox extends this file format such that a Bluetooth device can act as an input like a regular sensor or as an output. The latter allows sending sensor data from the phone to a peripheral, which opens up many additional applications not covered in this paper like controlling an Arduino by tilting the phone.

In the phyphox XML format a Bluetooth input defines the basic parameters to identify the correct device. Typically, this is either the advertised name or an advertised service. When opening an experiment configuration with a Bluetooth input, phyphox shows a scanning dialog offering nearby devices matching the name or service UUID.

The actual communication with the peripheral is then defined by one or more UUIDs of characteristics with a conversion function to define how the binary data should be interpreted. This can be a text format, a binary representation of a number or a more complex pattern in which several values are combined into one characteristic, for example when an accelerometer offers x, y and z axis in a single data packet. Additionally, the timestamps at which phyphox receives data can be recorded to create a time axis, but it should be noted that the time when the data was received may be delayed. On a busy connection this can easily introduce a jitter of several 100 ms, so if the peripheral is able to also submit a timestamp, this peripheral timestamp should be preferred.

The phyphox Bluetooth interface also allows to define data packages to configure the peripheral after connecting. For example, most Bluetooth sensor devices allow to set the data rate or sensitivity of a sensor through an additional characteristic. A specific configuration could set these as required for a specific experiment.

The resulting experiment configuration for a BLE peripheral is very versatile and so far, we were able to communicate with almost any commercially available BLE sensor. Incorporating BLE devices in this manner allows adding data analysis and visualization tools as well as combining data from external sensors with internal sensors as demonstrated exemplarily by Dorsel et al. (2022).

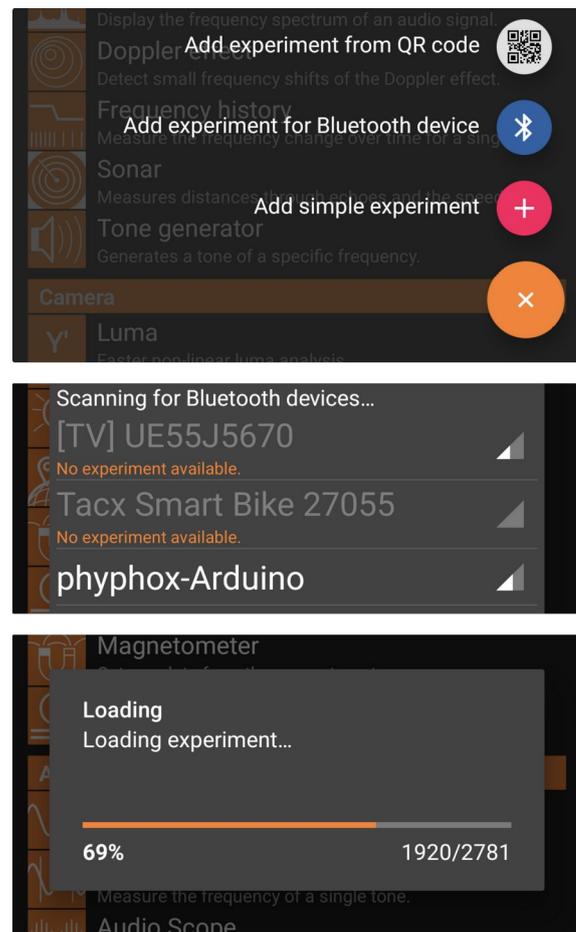

**Fig 2** Screenshots of phyphox (Android version) searching for and connecting to a Bluetooth device that implements the phyphox service. Top: The Bluetooth scan dialog can be reached via the plus button on the main screen. Middle: Compatible devices are highlighted in the scan dialog. Bottom: After selection the device transfers a suitable configuration.

The configuration for the device can be made available in phyphox in three ways. First, custom configurations are either sent directly to the phone as a file or shared as a QR code, which then again can be scanned from the main screen (see Fig. 2 top). This option is open for all users. Alternatively, the configuration is shipped with phyphox, if it is maintained by the phyphox team and allowed by the manufacturer. Supported devices then simply show up when scanning for Bluetooth devices from the plus button on the main screen. The third option is that the BLE device itself shares a suitable configuration with phyphox by implementing and advertising a specific phyphox service, thus making it available through the Bluetooth scan dialog (Fig. 2). As this can be highly technical and is usually only relevant to microcontroller projects. As a result, this third option is typically implemented through the support of the Arduino or MicroPython library as described below.

## 5. Difficulties of the generic BLE support

Using the generic BLE support in phyphox requires knowledge that cannot be expected from most science educators. Knowledge of the structure of BLE is not common and neither is a fundamental understanding of how numbers are represented in binary. These details are often not needed for a self-built device while finding out these details for a non-documented commercial device requires some experience. Similarly, setting up a BLE server can be very specific to the microcontroller that is being used and while these are usually properly documented, it is again knowledge that is not common among all Arduino or MicroPython enthusiasts.

While extensively documented, the XML file format for phyphox is not widely known. Requiring to learn this format is a massive hurdle for any quick and simple integration of custom built sensors and thus limits clearly the group of potential users. The same applies to submitting the configuration file to phyphox by the peripheral itself. In principle, a QR code would suffice, but having the device advertise and use the phyphox service is much more convenient to the user of the sensor (i.e. the students), but requires even more technical knowledge.

Considering these difficulties, it is not surprising that we have seen little use of the generic BLE support in phyphox except for specific requests that typically someone from the phyphox team then implemented. To overcome these difficulties, it was decided to encapsulate all these prerequisites into an Arduino library, allowing users with only basic knowledge to use all these features with only few lines of code in the familiar Arduino language without having to know anything about the inner workings or the file format of phyphox.

## 6. Arduino library

The Arduino library that allows to connect a BLE-enabled microcontroller to phyphox is primarily distributed through the Library Manager that is integrated into the Arduino IDE, meaning that users can simply search for "phyphox" in their known environment to find and install it. This also installs several examples to get the user started, one of which is the minimal example in Listing 1 that generates random numbers on the microcontroller and plots them in phyphox as shown in Figure 3.

```
#include <phyphoxBle.h>

void setup() {
  PhyphoxBLE::start(); //Start the BLE server
}

void loop() {
  float randomNumber = random(0,100);
  PhyphoxBLE::write(randomNumber); //Send value
  delay(50);
}
```

**Listing 1** Minimal example of using the phyphoxBle Arduino library to transfer and plot random numbers.

The massive improvement here is that on top of the default Arduino structure with a "loop" and a "setup" function along with the command to generate a random number and a short "delay", there are only three additional commands required to plot the data in phyphox. The user does not need to know anything about the XML configuration format or even how BLE works. One only needs an "include" statement to include the library, a call to "start" the BLE GATT server of the library, and a call to "write" the number to be plotted to phyphox.

Under the hood the library handles all the technical details that can not be expected to be known to an average Arduino enthusiast:

1. The library generates the XML configuration telling phyphox how to communicate with the Arduino and plot the data.

2. It sets up a GATT server using the right BLE library for this particular microcontroller.

3. It sets up services and characteristics to send (or receive) measurement data.

4. It sets up services and characteristics to send the XML configuration to phyphox without the need for a QR code.

5. It advertises itself with the phyphox-specific service UUID that allows phyphox to find it without having a matching configuration first.

6. It handles the binary encoding for the data transmission.

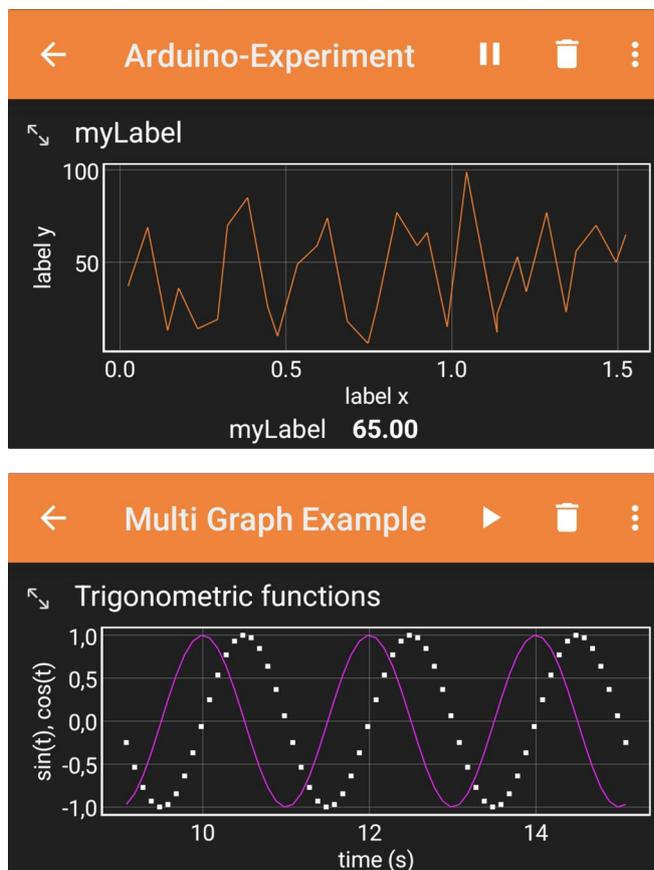

**Fig 3** Top: Result of the example given in Listing 1. Random numbers are generated on the microcontroller and plotted in phyphox. Bottom: Multi Graph Example included with the library. A sine function is generated on the microcontroller to illustrate different graph styles and plotting multiple values.

The result is that even users with little Arduino experience can add three lines of code to turn their own projects into devices that students can easily connect to with phyphox to plot sensor data over time on their smartphones or tablets. This can for example be the motion of an object in front of an ultrasound distance sensor over time, which can be realized by anyone who followed an online tutorial on how to read this sensor.

The downside of this minimal example is a lack of proper labeling as for example the graph axes are left with their defaults. But, the library exposes most of the phyphox XML format through a clean structure of objects representing different elements like graphs, simple numeric displays, buttons or text boxes as well as their properties like labels, colors and styles. It is also possible to plot multiple values in a single graph (Fig. 3 bottom) or map different values to different axes and therefore to plot a value of one sensor against a value of another sensor instead of representing a time series. Therefore, after the initial success of sending data with few lines of code, one can construct a custom user interface with labels and instructions for any DIY microcontroller based measuring instrument.

| Manufacturer | Models |
|---|---|
| Arduino | Nano 33 BLE<br>Nano 33 IoT<br>Nano Sense<br>Uno R4 Wifi |
| Espressif Systems | Any ESP 32-based board<br>Tested with ESP32-WROOM-32D, ESP32-DevKitS and ESP32-CAM |

*Table 1* List of microcontrollers currently supported by the phyphox Arduino library.

A notable usage example published shortly after the release of the library is the use of an Arduino Nano 33 BLE Sense with phyphox by Bouquet et al. (2022). This particular microcontroller has several sensors on board that were made available to students with detailed building instructions.

At the time of writing, the library supports several Arduino-branded boards (Arduino Nano 33 BLE, Arduino Nano 33 BLE Sense, Arduino Nano 33 IoT) as well as any ESP-32 based development board (see Table 1). Support for new microcontrollers will be continuously added.

## 7. MicroPython library

For self-made devices, Arduino is by far the most popular choice with the widest range of tutorials available and many examples for educational purposes (Goncalves et al., 2023). The process of designing a circuit to communicate with a particular sensor and, even more so, the process of writing code to read the sensor and maybe even analyze the data can also be an instructive and motivating project for students. However, while the Arduino language or rather its closely related archetypes C and C++ have had long been the most popular language for introductory programming courses in higher education (Aleksić & Ivanović, 2016), nowadays syntactically easy languages like Python have been found to facilitate student's learning of programming concepts (Mehmood et al., 2020). Therefore, Python can often be found in programming courses for science students as it also plays an important role in data analysis.

In this situation, using the Arduino library can lead to confusion for programming novices who face the differences in Arduino and Python syntax. To solve this issue, we also ported the library to MicroPython, which resulted in the minimal example in Python, shown in Listing 2.

```python
from phyphoxBLE import PhyphoxBLE
import time
import random

def main():
    p = PhyphoxBLE() #Instantiate PhyphoxBLE
    p.start()       #Start the BLE server

    while True:
        randomNumber = random.randint(0,100)
        p.write(randomNumber) #Send value
        time.sleep_ms(50)
```

**Listing 2** Minimal example of using the phyphoxBLE MicroPython module to transfer and plot random numbers.

This example is logically identical to the random numbers example in Arduino as shown in Listing 1. Besides the obviously different syntax, there are only minor language specific differences. The most obvious one is the import of additional modules right at the beginning, which is explicitly required in Python but done under the hood by the Arduino IDE. The other difference is that in this example the PhyphoxBLE class needs to be instantiated while the Arduino library works with static function calls.

## 8. Evaluation of the approach

In a lab course with 36 participants, students were given an ESP32-based development board, a bread board, some cables and an optical time-of-flight distance sensor. With a wiring scheme, a code snippet that reads a distance from the sensor and the examples from the MicroPython library, all students succeeded to create a distance-time plot in phyphox and used this as a basis to solve various practical tasks and measurements on a spring oscillator. This even applied to programming novices. In an accompanying questionnaire, the students assessed their previous experience with microcontrollers to be low (mean of 1.6 on a scale from 1 (for "no experience") to 4 (for "much experience")), but confirmed afterwards that they now felt confident to use the microcontroller from the experiment in other projects (mean of 3.1 on a scale from 1 (no) to 4 (yes), translation of original German question: "Are you confident to use the microcontroller in other projects?").

## 9. Conclusion

Integrating Bluetooth Low Energy sensors into phyphox opens up a range of additional experiments thanks to the huge selection of cheaply available sensors.

The first generic BLE support granted access to a selection of devices for which we either included a configuration in phyphox or shared it for specific requests. Adding more devices, however, is a highly technical task which is too inaccessible or time consuming for most science educators and learners. This has been alleviated by an Arduino library, allowing quick and easy phyphox integration into self-made sensor electronics for any science educator with basic interest in the Arduino platform and advanced learners who receive moderate support. The task of integrating phyphox became comparable to integrating one of the sensors themselves by simply loading and using the appropriate library from the Arduino IDE. With an additional MicroPython variant we have also added a version that can be more suitable to students who just start learning a programming language. The libraries open the door to an easy realization of rather complex experiments, using cheap sensors, microcontrollers and the creativity of educators and advanced learners.

## Source code

The state of the phyphox Arduino library at the time of submission of this paper has been archived under the DOI https://doi.org/10.5281/zenodo.10849073. The state of the MicroPython module has been archived under the DOI https://doi.org/10.5281/zenodo.10849061. Both have been released under the GNU Lesser General Public Licence v3.0 and the latest versions can be found under https://github.com/phyphox/phyphox-arduino and https://github.com/phyphox/phyphox-micropython, resepectivley.

## Acknowledgements

Parts of the BLE interface were developed in context of the project "Lehrerbildung Aachen" (LeBiAC2, FKZ 01JA1813) funded by the German Federal Ministry of Education and Research in their funding program "Qualitätsoffensive Lehrerbildung" (Phase 2). Further development was funded by the RWTH Aachen University's grant "Exploratory Teaching Space" (ETS).